\documentclass{article}
\usepackage{fullpage}
\usepackage[american]{babel}
\usepackage{amssymb}
\usepackage{amsmath}
\usepackage{algorithmic}
\usepackage{algorithm}
\usepackage{epsfig}
\usepackage{xcolor}
\usepackage{listings}

\lstdefinestyle{customc}{
  belowcaptionskip=1\baselineskip,
  breaklines=true,
  frame=L,
  xleftmargin=\parindent,
  language=C,
  showstringspaces=false,
  basicstyle=\footnotesize\ttfamily,
  keywordstyle=\bfseries\color{green!40!black},
  commentstyle=\itshape\color{purple!40!black},
  identifierstyle=\color{blue},
  stringstyle=\color{orange},
}

\newcommand{\eqn}[1]{(\ref{#1})}
\newcommand{\beql}[1]{\begin{equation}\label{#1}}
\newcommand{\eeq}{\end{equation}}
\newtheorem{theo}{Theorem}

\title{A note on the largest number of red nodes in red-black trees}
\author{Yingjie Wu, Daxin Zhu, Lei Wang and Xiaodong Wang}

\begin{document}
\maketitle

\begin{abstract}
In this paper, we are interested in the number of red nodes in red-black trees. We first present an $O(n^2\log n)$ time dynamic programming solution for computing $r(n)$, the largest number of red internal nodes in a red-black tree on $n$ keys.
Then the algorithm is improved to some $O(\log n)$ time recursive and nonrecursive algorithms. Based on these improved algorithms we finally find a closed-form solution of $r(n)$.
\end{abstract}

\section{Introduction}

A red-black tree is a special type of binary tree, used in computer science to organize pieces of comparable data, such as strings or numbers.
The original data structure was invented in 1972 by Rudolf Bayer\cite{2} with its name 'symmetric binary B-tree'. In a paper entitled 'A Dichromatic Framework for Balanced Trees', Guibas and Sedgewick named it red-black tree in 1978\cite{4}. In their paper they studied the properties of red-black trees at length and introduced the red/black color convention. Andersson \cite{1} gives a simpler-to-code variant of red-black trees. Weiss \cite{6} calls this variant AA-trees. An AA-tree is similar to a red-black tree except that left children may never be red.
In 2008, Sedgewick introduced a simpler version of the red-black tree called the left-leaning red-black tree\cite{5} by eliminating a previously unspecified degree of freedom in the implementation. Red-black trees can be made isometric to either 2-3 trees or 2-4 trees,\cite{5} for any sequence of operations.

A red-black tree is a binary search tree with one extra bit of storage per node: its color, which can be either red or black. It satisfies the following red-black properties\cite{5}:

(1) A node is either red or black.

(2) The root is black.

(3) All leaves (NIL) are black.

(4) Every red node must have two black child nodes.

(5) Every path from a given node to any of its descendant leaves contains the same number of black nodes.

The number of black nodes on any simple path from, but not including, a node $x$ down to a leaf is called the black-height of the node, denoted $bh(x)$. By the property (5),the notion of black-height is well defined, since all descending simple paths from the node have the same number of black nodes. The black-height of a red-black tree is defined to be the black-height of its root.

The property (2) is sometimes omitted in practice. Since the root can always be changed from red to black, but not necessarily vice-versa, this property has little effect on analysis. A binary search tree that satisfies red-black properties (1), (3), (4), and (5) is sometimes called a relaxed red-black tree. In this paper we will discuss the relaxed red-black tree and call a relaxed red-black tree a red-black tree.

We are interested in the number of red nodes in red-black trees in this paper. We will investigate the problem that in a red-black tree on $n$ keys, what is the largest possible ratio of red internal nodes to black internal nodes.

\section{A dynamic programming algorithm}

Let $T$ be a red-black tree on $n$ keys.
The largest and the smallest number of red internal nodes in a red-black tree on $n$ keys can be denoted as $r(n)$ and $s(n)$ respectively.
The values of $r(n)$ and $s(n)$ can be easily observed for the special case of $n=2^k-1$ .
It is obvious that in this case, when the node colors are alternately red and black from the bottom level to the top level of $T$, the number of red internal nodes of $T$ must be maximal. When all of its internal nodes are black, the number of red internal nodes of $T$ must be minimal.
Therefore, in the case of $n=2^k-1$, we have,

\[
\begin{array}{l}
r(n)=r(2^{k} -1)=\displaystyle{
\sum _{i=0}^{\left\lfloor (k-1)/2\right\rfloor }2^{k-2i-1}}
=2^{k-1} \displaystyle{\sum _{i=0}^{\left\lfloor (k-1)/2\right\rfloor }\frac{1}{4^{i}}}
\\
=\displaystyle{\frac{2^{k-1} }{3} \left(4-\frac{1}{4^{\left\lfloor (k-1)/2\right\rfloor}}\right)}
=\displaystyle{\frac{2^{k+1}-2^{k-1-2\left\lfloor (k-1)/2\right\rfloor } }{3}}\\
=\displaystyle{\frac{2^{k+1} -2^{(k-1)\texttt{ mod }2} }{3}}=\displaystyle{\frac{2^{k+1} -2+k \texttt{ mod } 2}{3}}\\
=\displaystyle{\frac{2(2^{k} -1)+k\texttt{ mod }2}{3}}=\displaystyle{\frac{2n+\log (n+1)\texttt{ mod }2}{3}}\\
\end{array}
\]

In the general cases, we denote the largest number of red internal nodes in a subtree of size $i$ and black-height $j$ to be $a(i,j,0)$ when its root red and $a(i,j,1)$ when its root black respectively. Since in a red-black tree on $n$ keys we have $\frac{1}{2}\log n\leq j\leq 2\log n$, we have,

\beql{eq21}
\gamma(n,k)=\max\limits_{\frac{1}{2}\log n\leq j\leq 2\log n}a(n,j,k)
\eeq

Furthermore, for any $1\leq i\leq n, \frac{1}{2}\log i\leq j\leq 2\log i$, we can denote,

\beql{eq22}
\left\{\begin{array}{l}
\alpha_1(i,j)=\max\limits_{0\leq t\leq i/2 }\{a(t,j-1,1)+a(i-t-1,j-1,1)\}\\
\alpha_2(i,j)=\max\limits_{0\leq t\leq i/2 }\{a(t,j,0)+a(i-t-1,j,0)\}\\
\alpha_3(i,j)=\max\limits_{0\leq t\leq i/2 }\{a(t,j-1,1)+a(i-t-1,j,0)\}\\
\alpha_4(i,j)=\max\limits_{0\leq t\leq i/2 }\{a(t,j,0)+a(i-t-1,j-1,1)\}\\
\end{array} \right.
\eeq

\begin{theo}\label{th1}

For each $1\leq i\leq n, \frac{1}{2}\log i\leq j\leq 2\log i$, the values of $a(i,j,0)$ and $a(i,j,1)$ can be computed by the following dynamic programming formula.
\end{theo}

\beql{eq23}
\left\{\begin{array}{l}
a(i,j,0)=1+\alpha_1(i,j)\\
a(i,j,1)=\max\{\alpha_1(i,j),\alpha_2(i,j),\alpha_3(i,j)\}\\
\end{array} \right.
\eeq

\noindent{\bf Proof.}

For each $1\leq i\leq n, \frac{1}{2}\log i\leq j\leq 2\log i$, let $T(i,j,0)$ be a red-black tree on $i$ keys and black-height $j$ with the largest number of red internal nodes, when its root red. $T(i,j,1)$ can be defined similarly when its root black.
The red internal nodes of $T(i,j,0)$ and $T(i,j,1)$ must be $a(i,j,0)$ and $a(i,j,1)$ respectively.

(1) We first look at $T(i,j,0)$. Since its root is red, its two sons must be black, and thus the black-height of the corresponding subtrees $L$ and $R$ must be both $j-1$.
For each $0\leq t\leq i/2$, subtrees $T(t,j-1,1)$ and $T(i-t-1,j-1,1)$ connected to a red node will be a red-black tree on $i$ keys and black-height $j$. Its number of red internal nodes must be $1+a(t,j-1,1)+a(i-t-1,j-1,1)$. In such trees, $T(i,j,0)$ achieves the maximal number of red internal nodes. Therefore, we have,
\beql{eq24}
a(i,j,0)\geq\max\limits_{0\leq t\leq i/2 }\{1+a(t,j-1,1)+a(i-t-1,j-1,1)\}
\eeq

On the other hand, we can assume the sizes of subtrees $L$ and $R$ are $t$ and $i-t-1$, $0\leq t\leq i/2$, WLOG. If we denote the number of red internal nodes in $L$ and $R$ to be $r(L)$ and $r(R)$, then we have that $r(L)\leq a(t,j-1,1)$ and $r(R)\leq a(i-t-1,j-1,1)$. Thus we have,
\beql{eq25}
a(i,j,0)\leq 1+\max\limits_{0\leq t\leq i/2 }\{a(t,j-1,1)+a(i-t-1,j-1,1)\}
\eeq

Combining \eqn{eq24} and \eqn{eq25}, we obtain,
\beql{eq26}
a(i,j,0)=1+\max\limits_{0\leq t\leq i/2 }\{a(t,j-1,1)+a(i-t-1,j-1,1)\}
\eeq

(2) We now look at $T(i,j,1)$. Since its root is black, there can be 4 cases of its two sons such as red and red, black and black, black and red or red and black. If the subtree $L$ or $R$ has a red root, then the black-height of the corresponding subtree must be $j$, otherwise, if its root is black, then the black-height of the subtree must be $j-1$.

In the first case, both of the subtrees $L$ and $R$ have a black root. For each $0\leq t\leq i/2$, subtrees $T(t,j-1,1)$ and $T(i-t-1,j-1,1)$ connected to a black node will be a red-black tree on $i$ keys and black-height $j$. Its number of red internal nodes must be $a(t,j-1,1)+a(i-t-1,j-1,1)$. In such trees, $T(i,j,1)$ achieves the maximal number of red internal nodes. Therefore, we have,
\beql{eq27}
a(i,j,1)\geq\max\limits_{0\leq t\leq i/2 }\{a(t,j-1,1)+a(i-t-1,j-1,1)\}=\alpha_1(i,j)
\eeq

For the other three cases, we can conclude similarly that

\beql{eq28}
a(i,j,1)\geq\max\limits_{0\leq t\leq i/2 }\{a(t,j,0)+a(i-t-1,j,0)\}=\alpha_2(i,j)
\eeq
\beql{eq29}
a(i,j,1)\geq\max\limits_{0\leq t\leq i/2 }\{a(t,j-1,1)+a(i-t-1,j,0)\}=\alpha_3(i,j)
\eeq
\beql{eq210}
a(i,j,1)\geq\max\limits_{0\leq t\leq i/2 }\{a(t,j,0)+a(i-t-1,j-1,1)\}=\alpha_4(i,j)
\eeq

Therefore, we have,
\beql{eq211}
a(i,j,1)\geq\max\{\alpha_1(i,j),\alpha_2(i,j),\alpha_3(i,j),\alpha_4(i,j)\}
\eeq

On the other hand, we can assume the sizes of subtrees $L$ and $R$ are $t$ and $i-t-1$, $0\leq t\leq i/2$, WLOG. In the first case, if we denote the number of red internal nodes in $L$ and $R$ to be $r(L)$ and $r(R)$, then we have that $r(L)\leq a(t,j-1,1)$ and $r(R)\leq a(i-t-1,j-1,1)$, and thus we have,
\beql{eq212}
a(i,j,1)\leq \max\limits_{0\leq t\leq i/2 }\{a(t,j-1,1)+a(i-t-1,j-1,1)\}=\alpha_1(i,j)
\eeq

For the other three cases, we can conclude similarly that

\beql{eq213}
a(i,j,1)\leq\max\limits_{0\leq t\leq i/2 }\{a(t,j,0)+a(i-t-1,j,0)\}=\alpha_2(i,j)
\eeq
\beql{eq214}
a(i,j,1)\leq\max\limits_{0\leq t\leq i/2 }\{a(t,j-1,1)+a(i-t-1,j,0)\}=\alpha_3(i,j)
\eeq
\beql{eq215}
a(i,j,1)\leq\max\limits_{0\leq t\leq i/2 }\{a(t,j,0)+a(i-t-1,j-1,1)\}=\alpha_4(i,j)
\eeq

Therefore, we have,
\beql{eq216}
a(i,j,1)\leq\max\{\alpha_1(i,j),\alpha_2(i,j),\alpha_3(i,j),\alpha_4(i,j)\}
\eeq

Combining \eqn{eq211} and \eqn{eq216}, we obtain,
\beql{eq217}
a(i,j,1)=\max\{\alpha_1(i,j),\alpha_2(i,j),\alpha_3(i,j),\alpha_4(i,j)\}
\eeq

It is readily seen by the symmetry of the case 3 and case 4 that $\alpha_3(i,j)=\alpha_4(i,j)$, for each $1\leq i\leq n, \frac{1}{2}\log i\leq j\leq 2\log i$, and finally we have,

\beql{eq218}
a(i,j,1)=\max\{\alpha_1(i,j),\alpha_2(i,j),\alpha_3(i,j)\}
\eeq

The proof is complete.
$\blacksquare$

According to Theorem \ref{th1}, our algorithm for computing $a(i,j,k)$ is a standard 2-dimensional dynamic programming algorithm. It is obvious that the algorithm requires $O(n^2\log n)$ time and $O(n\log n)$ space.

\section{The improved dynamic programming solutions}

We have computed $r(n)$ and the corresponding red-black trees using the dynamic programming algorithm algorithm.
Some pictures of the computed red-black trees with largest number of red nodes are listed in Fig. 1.
From these pictures of the red-black trees with largest number of red nodes in various size, we can observe some properties of $r(n)$ and the corresponding red-black trees as follows.

\begin{figure}
\centering
\includegraphics[width=14cm,height=9cm]{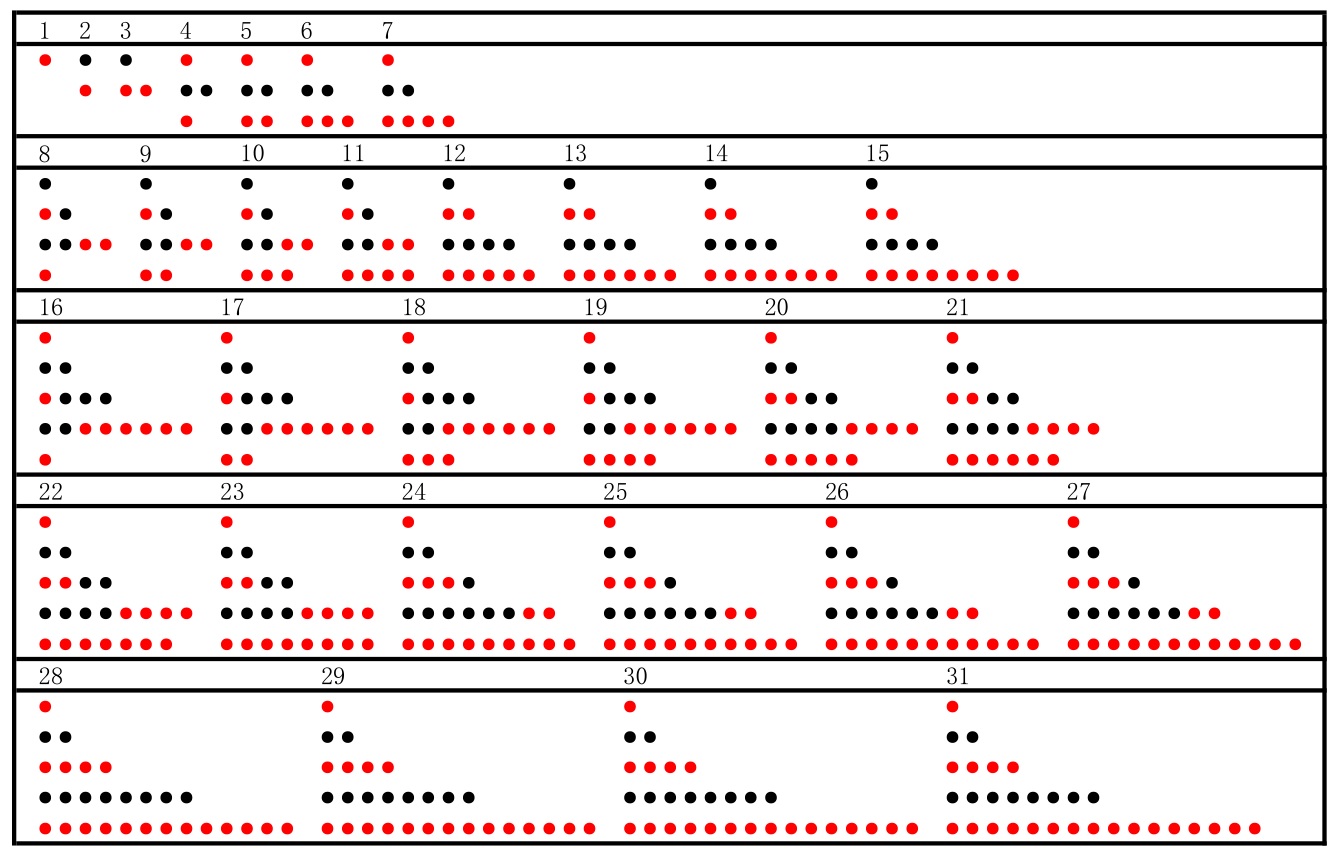}
\caption{The maximal red-black trees}
\end{figure}

(1) The red-black tree on $n$ keys with $r(n)$ red nodes can be realized in a complete binary search tree, called a maximal red-black tree.

(2) In a maximal red-black tree, the colors of the nodes on the left spine are alternatively red, black, $\cdots$, from the bottom to the top, and thus the black-height of the red-black tree must be $\frac{1}{2}\log n$.

(3) In a maximal red-black tree of $k$ levels, if all of the nodes of the last two levels ($k, k-1$) and all of the black nodes of the last third level ($k-2$) are removed, the remaining tree is also a maximal red-black tree.

From these observations, we can improve the dynamic programming formula of Theorem 1 further.
By the observation (3), the time complexity of the algorithm can be reduced substantially as follows.

\begin{theo}\label{th2}

Let $n$ be the number of keys in a red-black tree, and $r(n)$ be the largest number of red nodes in a red-black tree on $n$ keys. The values of $r(n)$ can be computed by the following recursive formula.

\beql{eq31}
r(n)=\left\{\begin{array}{ll}
n-\lfloor\log n\rfloor & n<8\\
r(p)+q & n\geq 8\\
\end{array} \right.
\eeq

where

\beql{eq32}
\left\{\begin{array}{l}
p=2^{\lfloor\log n\rfloor-2}+\lceil(n-2^{\lfloor\log n\rfloor}+1)/4\rceil-1\\
q=n-2^{\lfloor\log n\rfloor-1}-2\lceil(n-2^{\lfloor\log n\rfloor}+1)/4\rceil+1 \\
\end{array} \right.
\eeq
\end{theo}

\noindent{\bf Proof.}

Let $T$ be a maximal red-black tree of size $n$. It is obvious that $T$ has $k=1+\lfloor\log n\rfloor$ levels.

(1) The formula can be verified directly for the case of $n<8$.

(2) In the case of $n\geq 8$, we have $k>3$. The number of nodes in the last level of $T$ must be $s=n-2^{\lfloor\log n\rfloor}+1$. These nodes are all red nodes of $T$. It is readily seen that every 4 red nodes in the last level correspond to 2 black nodes in level $k-1$ of $T$.
Thus the number of black nodes in level $k-1$ must be $b=2\lceil(n-2^{\lfloor\log n\rfloor}+1)/4\rceil$. It follows that the number of red nodes in level $k-1$ of $T$ is $2^{\lfloor\log n\rfloor-1}-b$.
Therefore, the number of red nodes in the last two levels of $T$ is $s+2^{\lfloor\log n\rfloor-1}-b$,
which is exactly $q=n-2^{\lfloor\log n\rfloor-1}-2\lceil(n-2^{\lfloor\log n\rfloor}+1)/4\rceil+1$.

Let $T'$ be the subtree of $T$ by removing all of the nodes of the last two levels ($k, k-1$) and all of the black nodes in level $k-2$ from $T$.
Since every 2 black nodes in level $k-1$ correspond to 1 red node in level $k-2$ of $T$, the number of red nodes in level $k-2$ is obviously $b/2$, and thus the size of $T'$ must be $2^{\lfloor\log n\rfloor-2}+b/2$,
which is exactly $p=2^{\lfloor\log n\rfloor-2}+\lceil(n-2^{\lfloor\log n\rfloor}+1)/4\rceil-1$.
It follows from observation (3) that $r(n)=r(p)+q$.

The proof is complete.
$\blacksquare$

According to Theorem \ref{th2}, a new recursive algorithm for computing the largest number of red internal nodes in a red-black tree on $n$ keys can be implemented in $O(\log n)$ time.

For the same problem, we can build another efficient algorithm in a different point of view.
If we list the sequence of the values of $r(n)$ as a triangle $t(i,j),i=0,1,\cdots, j=1,2,\cdots, 2^i$ as shown in Table 1, then we can observe some interesting structural properties of $r(n)$.

\begin{table}
\caption{A triangle of sequence $r(n)$}
\begin{center}
\begin{tabular}{c|cccccccccccccccc}
\hline\hline
Row No. &&&&&&&&$r(n)$&&&&&&&\\\hline
0&0&&&&&&&&&&&&&&&\\\hline
1&1&1&&&&&&&&&&&&&&\\\hline
2&2&2&\textcolor[rgb]{1.00,0.00,0.00}{\textbf{3}}&4&&&&&&&&&&&&\\\hline
3&\textcolor[rgb]{1.00,0.00,0.00}{\textbf{5}}&4&5&6&7&7&8&9&&&&&&&&\\\hline
4&10&9&10&11&12&12&13&14&\textcolor[rgb]{1.00,0.00,0.00}{\textbf{15}}&15&16&17&18&18&19&20\\\hline
5&\textcolor[rgb]{1.00,0.00,0.00}{\textbf{21}}&19&20&21&22&22&23&24&25&25&26&27&28&28&29&30\\\hline
\end{tabular}
\end{center}
\end{table}

It is readily seen from Table 2 that the values in each row have some regular patterns as follows.

(1) For the fist elements $t(i,1)$ in each row $i=0,1,\cdots$, we have,
$t(2j+1,1)=2t(2j,1)+1,t(2j,1)=2t(2j-1,1),j=1,2,\cdots$.

(2) For the elements $t(i,2^{i-1}+1)$ in each row $i=1,2,\cdots$, we have, $t(i,2^{i-1}+1)=2^i-1$.

(3) For the elements $t(i,j), 2\leq j\leq 2^{i-1}$ in each row $i=2,3, \cdots$, we have, $t(i,j)=t(i-1,j)+c$ where $c$ is a constant.

(4) For the elements $t(i,j), 2^{i-1}+2\leq j\leq 2^i$ in each row $i=2,3, \cdots$, we have, $t(i,j)=t(i-1,j-2^{i-1})+d$ where $d$ is a constant.

In the insight of these observations, we can build another efficient algorithm to compute $r(n)$ as follows.

\begin{theo}\label{th3}

Let $n$ be the number of keys in a red-black tree, and $r(n)$ be the largest number of red nodes in a red-black tree on $n$ keys. If the values of $r(n)$ are listed as a triangle $t(i,j),i=0,1,\cdots, j=1,2,\cdots, 2^i$ as shown in Table 2, then the values of $t(i,j)$ can be computed by the following recursive formula.

\beql{eq33}
t(i,j)=\left\{\begin{array}{ll}
i & i<2\\
\xi(i) & 2\leq i, j=1\\
t(i-1,j)+\xi(i-1) & 2\leq i, 2\leq j\leq2^{i-1}\\
2^i-1 & 2\leq i, j=2^{i-1}+1\\
t(i-1,j-2^{i-1})+\eta(i) & 2\leq i, 2^{i-1}+2\leq j\leq2^i\\
\end{array} \right.
\eeq

where

\beql{eq34}
\left\{\begin{array}{l}
\xi(i)=\frac{2}{3}\left(2^i-\frac{3+(-1)^i}{4}\right)=\lceil\frac{2}{3}(2^i-1)\rceil\\
\eta(i)=\frac{1}{3}(2^{i+1}+(-1)^i)=\lfloor\frac{1}{3}(2^{i+1}+1)\rfloor\\
\end{array} \right.
\eeq
\end{theo}

\noindent{\bf Proof.}

(1) The formula can be verified directly for the case of $i<2$.

(2) In the case of $i\geq 2$, we will consider the following 4 cases.

(2.1) In the case of $2\leq i, j=1$, we have,

\beql{eq35}
t(i,1)=\left\{\begin{array}{ll}
2t(i-1,1)+1 & i \text{  odd}\\
2t(i-1,1) & i \text{  even}\\
\end{array} \right.
\eeq

Therefore, if we denote $\xi(i)=t(i,1)$, then in general we have,

$\xi(i)=2\xi(i-1)+\frac{1-(-1)^i}{2}$

$=4\xi(i-2)+2\frac{1-(-1)^{i-1}}{2}+\frac{1-(-1)^i}{2}$

$=\cdots$

$=2^{i-1}+\displaystyle{\sum_{j=0}^{i-2}2^j\frac{1-(-1)^{i-j}}{2}}$

$=2^{i-1}+\displaystyle{\sum_{j=0}^{i-2}2^{j-1}+\sum_{j=0}^{i-2}(-1)^{i-j}2^{j-1}}$

$=2^{i-1}+2^{i-2}-\frac{1}{2}-\displaystyle{(-1)^i\sum_{j=0}^{i-2}(-1)^{j-1}2^{j-1}}$

$=3\cdot2^{i-2}-\frac{1}{2}-\frac{1}{6}\left\{\begin{array}{ll}
2\cdot 4^{\frac{i-2}{2}} +1& i \text{  even }\\
4^{\lceil\frac{i-2}{2}\rceil} -1 & i \text{  odd}\\
\end{array} \right.
$

$=3\cdot2^{i-2}-\frac{1}{2}-\frac{1}{6}\left\{\begin{array}{ll}
2\cdot 4^{\frac{i-2}{2}} +1& i \text{  even}\\
4^{\lceil\frac{i-2}{2}\rceil} -1 & i \text{  odd}\\
\end{array} \right.
$

$=3\cdot2^{i-2}-\frac{1}{3}2^{i-2}-\left\{\begin{array}{ll}
\frac{2}{3}& i \text{  even}\\
\frac{1}{3} & i \text{  odd}\\
\end{array} \right.
$

$=\frac{1}{3}\cdot2^{i+1}-\frac{1}{3}-\frac{1}{3}\left(\frac{1+(-1)^i}{2}\right)$

$=\frac{2}{3}\left(2^i-\frac{3+(-1)^i}{4}\right)=\lceil\frac{2}{3}(2^i-1)\rceil$

(2.2) In the case of $2\leq i, 2\leq j\leq2^{i-1}$, we have, $t(i,j)=t(i-1,j)+c$ where $c$ is a constant for all $j$.
It can be seen from Table 2 that in this case $c=\xi(i-1)$.

(2.3) In the case of $2\leq i, j=2^{i-1}+1$, it is obvious that $t(i,2^{i-1}+1)=2^i-1$.

(2.4) In the case of $2\leq i, 2^{i-1}+2\leq j\leq2^i$, we have, $t(i,j)=t(i-1,j-2^{i-1})+\eta(i)$ where $\eta(i)$ is a constant for all $j$.
It can be seen from Table 2 that in this case the constant $\eta(i)$ is different from $i$ is odd and even.
When $i$ is even, we have $\eta(i)=2^{i}-\xi(i-1)$. If $i$ is odd, then we have $\eta(i)=2^{i}-\xi(i-1)-1$.
Therefore, in general we have,

$\displaystyle{\eta(i)=2^i-\xi(i-1)-\frac{1-(-1)^i}{2}}$

$\displaystyle{=2^{i}-\frac{1}{3}\left(2^{i}-\frac{3+(-1)^{i-1}}{2}\right)-\frac{1-(-1)^i}{2}}$

$\displaystyle{=\frac{1}{3}2^{i+1}+\frac{3-(-1)^i-3+3(-1)^i}{6}}$

$\displaystyle{=\frac{1}{3}(2^{i+1}+(-1)^i)=\lfloor\frac{1}{3}(2^{i+1}+1)\rfloor}$

The proof is complete.
$\blacksquare$

\section{The closed-form solution of $r(n)$}

For any positive integer $n$, let its binary expansion be $n=\displaystyle{\sum_{i=0}^{\lfloor\log n\rfloor}b_i2^i}$. The binary weight $a(n)$ of $n$ is defined to be the number of 1's in its binary expansion, i.e.
\beql{eq36}
a(n)=\displaystyle{\sum_{i=0}^{\lfloor\log n\rfloor}b_i}
\eeq

Two other corresponding functions, the even binary weight $e(n)$ and the odd binary weight $o(n)$ are defined as
\beql{eq37}
e(n)=\displaystyle{\sum_{i=0}^{\lceil(\lfloor\log n\rfloor-1)/2\rceil} b_{2i}}
\eeq

and

\beql{eq38}
o(n)=\displaystyle{\sum_{i=0}^{\lfloor(\lfloor\log n\rfloor-1)/2\rfloor} b_{2i+1}}
\eeq

It is obvious that $a(n)=e(n)+o(n)$.

\begin{theo}\label{th4}

Let $n$ be the number of keys in a red-black tree, and its binary expression be $n=\displaystyle{\sum_{i=0}^{\log n}b_i2^i}$, then $r(n)$ can be computed by the following formula,

\beql{eq39}
r(n)=\frac{1}{3}(2n+a(n)+o(n))-\left\lfloor\frac{\lfloor\log n\rfloor+1}{2}\right\rfloor
\eeq

where $a(n)$ and $o(n)$ are the binary weight and the odd binary weight of $n$  respectively.

\end{theo}

\noindent{\bf Proof.}

Let the binary expansion of $n$ be $n=\displaystyle{\sum_{i=0}^{\lfloor\log n\rfloor}b_i2^i}$.
We can conclude from Algorithm 4 that

$\displaystyle{r(n)=1+\sum_{i=1}^{\lfloor\log n\rfloor}(\eta(i)b_{i-1}+\xi(i-1)(1-b_{i-1}))}$

$\displaystyle{=1+\sum_{i=1}^{\lfloor\log n\rfloor}((\eta(i)-\xi(i-1))b_{i-1}+\xi(i-1))}$

From formula \eqn{eq34} we know

$$\displaystyle{\xi(i-1)=\frac{2}{3}\left(2^{i-1}-\frac{3+(-1)^{i-1}}{4}\right)}$$

$$\displaystyle{\eta(i)-\xi(i-1)=\frac{1}{3}(2^{i+1}+(-1)^i)-\frac{2}{3}\left(2^{i-1}-\frac{3+(-1)^{i-1}}{4}\right)=\frac{1}{3}\left(2^{i}+\frac{3+(-1)^{i}}{2}\right)}$$

Therefore, we have

$$\displaystyle{r(n)=1+\frac{1}{3}\sum_{i=1}^{\lfloor\log n\rfloor}\left(2^{i}+\frac{3+(-1)^{i}}{2}\right)b_{i-1}+\frac{1}{3}\sum_{i=1}^{\lfloor\log n\rfloor}\left(2^{i}-\frac{3-(-1)^{i}}{2}\right)}$$

If we denote

$$r_1(n)=\displaystyle{\frac{1}{3}\sum_{i=1}^{\lfloor\log n\rfloor}\left(2^{i}+\frac{3+(-1)^{i}}{2}\right)b_{i-1}}$$

$$r_2(n)=\displaystyle{\frac{1}{3}\sum_{i=1}^{\lfloor\log n\rfloor}\left(2^{i}-\frac{3-(-1)^{i}}{2}\right)}$$

then we have

$$r(n)=1+r_1(n)+r_2(n)$$

Furthermore, we have

$r_1(n)=\displaystyle{\frac{1}{3}\sum_{i=1}^{\lfloor\log n\rfloor}\left(2^{i}+\frac{3+(-1)^{i}}{2}\right)b_{i-1}}$

$=\displaystyle{\frac{1}{3}\sum_{i=0}^{\lfloor\log n\rfloor-1}\left(2^{i+1}+\frac{3-(-1)^{i}}{2}\right)b_{i}}$

$=\displaystyle{\frac{2}{3}\sum_{i=0}^{\lfloor\log n\rfloor-1}b_{i}2^{i}+
\frac{1}{2}\sum_{i=0}^{\lfloor\log n\rfloor-1}b_{i}-
\frac{1}{6}\sum_{i=0}^{\lfloor\log n\rfloor-1}(-1)^ib_{i}}$

$=\displaystyle{\frac{2}{3}\left(n-2^{\lfloor\log n\rfloor}\right)+
\frac{1}{2}\left(a(n)-1\right)-
\frac{1}{6}\left(e(n)-o(n)-(-1)^{\lfloor\log n\rfloor}\right)}$

On the other hand,

$r_2(n)=\displaystyle{\frac{1}{3}\sum_{i=1}^{\lfloor\log n\rfloor}\left(2^{i}-\frac{3-(-1)^{i}}{2}\right)}$

$=\displaystyle{\frac{1}{3}\sum_{i=1}^{\lfloor\log n\rfloor}2^{i}-
\frac{1}{2}\sum_{i=1}^{\lfloor\log n\rfloor}1+
\frac{1}{6}\sum_{i=1}^{\lfloor\log n\rfloor}(-1)^{i}}$

$=\displaystyle{\frac{1}{3}\left(2^{\lfloor\log n\rfloor+1}-2\right)-
\frac{1}{2}\lfloor\log n\rfloor+
\frac{1}{6}\left(\frac{(-1)^{\lfloor\log n\rfloor}-1}{2}\right)}$

Summing up, we then have

$r(n)=1+r_1(n)+r_2(n)$

$=1+\displaystyle{\frac{2}{3}\left(n-2^{\lfloor\log n\rfloor}\right)+
\frac{1}{2}\left(a(n)-1\right)-
\frac{1}{6}\left(e(n)-o(n)-(-1)^{\lfloor\log n\rfloor}\right)}$

$+\displaystyle{\frac{1}{3}\left(2^{\lfloor\log n\rfloor+1}-2\right)-
\frac{1}{2}\lfloor\log n\rfloor+
\frac{1}{6}\left(\frac{(-1)^{\lfloor\log n\rfloor}-1}{2}\right)}$

$=\displaystyle{\frac{1}{3}\left(2n+a(n)+o(n)\right)-
\frac{1}{2}\lfloor\log n\rfloor+
\frac{1}{4}(-1)^{\lfloor\log n\rfloor}
+\left(1-\frac{1}{2}-\frac{2}{3}-\frac{1}{12}\right)}$

$=\displaystyle{\frac{1}{3}\left(2n+a(n)+o(n)\right)-
\frac{1}{4}\left(2\lfloor\log n\rfloor+1-(-1)^{\lfloor\log n\rfloor}\right)}$

$=\frac{1}{3}(2n+a(n)+o(n))-\left\lfloor\frac{\lfloor\log n\rfloor+1}{2}\right\rfloor$

The proof is complete.
$\blacksquare$

\section{Concluding remarks}
We have suggested a dynamic programming solution for computing $r(n)$, the largest number of red internal nodes in a red-black tree on $n$ keys.
The dynamic programming algorithm requires $O(n^2\log n)$ time and $O(n\log n)$ space. We then improve the algorithm to some $O(\log n)$ time recursive and nonrecursive algorithms. Based on these improved algorithms we finally come to a closed-form solution of $r(n)$.

The smallest number of red internal nodes in a red-black tree on $n$ keys can be computed analogously.

%\section*{Acknowledgment}
%This work was supported by the Natural Science Foundation of Fujian under Grant No.2013J01247, Fujian Provincial Key Laboratory of Data-Intensive Computing and Fujian University Laboratory of Intelligent Computing and Information Processing.

\end{document}